\documentclass[aps,
prd,
english,
nofootinbib,
reprint]{revtex4-1}

\usepackage{amsfonts,amsmath,amssymb}
\usepackage{graphicx}
\usepackage[utf8]{inputenc}
\usepackage{hyperref}
\usepackage{babel}
\usepackage{enumitem}
\newcommand\e{{\rm e}}

\newcommand\be{\begin{equation}}
\newcommand\ee{\end{equation}}
\newcommand\bea{\begin{eqnarray}}
\newcommand\eea{\end{eqnarray}}

\begin{document}

\def\rhoo{\rho_{_0}\!} %%neater subscript for rho, the disc level density function.
\def\rhooo{\rho_{_{0,0}}\!} %%neater subscript for rho, the disc level density function.

\begin{flushright}
\phantom{
{\tt arXiv:2006.$\_\_\_\_$}
}
\end{flushright}

%{\flushleft\vskip-1.4cm\vbox{\includegraphics[width=1.15in]{trumpet.pdf}}}

\title{%{\vskip-0.4cm}
A Non-Perturbative Random Matrix Model of ${\cal N}=2$ JT Supergravity}
\author{Clifford V. Johnson}
\email{cliffordjohnson@ucsb.edu Some of the work of this paper was carried out while the author was affiliated with the Department of Physics and Astronomy, University of Southern California, 
Los Angeles, CA 90089, U.S.A.}

\affiliation{Department of Physics, Broida Hall,   University of California, 
Santa Barbara, CA 93106, U.S.A.}
%\affiliation{\medskip Department of Physics and Astronomy, University of Southern California, Los Angeles, CA 90089, U.S.A.$^*$}

%\pacs{05.70.Ce,05.70.Fh,04.70.Dy}

\begin{abstract}
It is shown how to non-perturbatively define a random matrix model that captures key physics of ${\cal N}{=}2$  Jackiw-Teitelboim (JT) supergravity, going  well beyond the perturbative topological expansion defined recently by Turiaci and Witten. A decomposition into an infinite family of certain multicritical models is derived, leading to the definition of a non-linear ordinary differential equation from which the physics may be computed. Bogomol'nyi–Prasad–Sommerfield (BPS) states are naturally described by the model. The non-perturbative completions of the spectral densities for  non-BPS multiplets are readily extracted.

\end{abstract}


\maketitle

%\section{Introduction}

%\label{sec:introduction}
\section{Introduction}

Recent years have seen renewed excitement about the use of random matrix models to capture properties of two-dimensional quantum gravity. The new activity began with  the  Jackiw-Teitelboim~\cite{Jackiw:1984je,*Teitelboim:1983ux} (JT), model of gravity coupled to a scalar,  which is of considerable interest in large part because it (and some variants) arises as the near-horizon low temperature physics of a wide class of higher dimensional black holes. Understanding it  has yielded valuable insights into the nature of  quantum gravity.  Saad, Shenker and Stanford~\cite{Saad:2019lba}  demonstrated explicitly that there is a {\it perturbative} equivalence between  the Euclidean path integral quantization of JT gravity and  computations in a double-scaled~\cite{Brezin:1990rb,*Douglas:1990ve,*Gross:1990vs,*Gross:1990aw} random matrix model. (Here perturbation theory is topologically organized; contributions from surfaces of Euler characteristic~$\chi$ are weighted by $\e^{S_0\chi}$, where $S_0$ is the $T{=}0$ entropy.) Soon after, Stanford and Witten~\cite{Stanford:2019vob} showed that the equivalence extends to a wider class of models  including various ${\cal N}{=}1$ supersymmetric extensions. 

 It is extremely important to understand the non-perturbative definitions of these perturbative demonstrations, because there are key physical questions  that cannot be addressed  in perturbation theory. Such questions often involve access to properties of the  discrete spectrum that underlies the quantum gravity theory (as implied {\it e.g.,} by the finite Bekenstein-Hawking~\cite{Bekenstein:1973ur,*Hawking:1976de} entropy). Moreover, direct access to the discrete spectrum is key to understanding the nature of holographic duals in 2D. A program of formulating the matrix model non-perturbatively was begun in ref.~\cite{Johnson:2019eik} with ordinary JT and extended to various ${\cal N}{=}1$ variants in refs.~\cite{Johnson:2020heh,Johnson:2020exp,Johnson:2020mwi,Johnson:2021owr}~\footnote{The techniques have even been applied to certain models of flat space gravity and supergravity.~\cite{Kar:2022vqy,*Kar:2022sdc,*Rosso:2022tsv}. Note that other non-perturbative approaches to the random matrix model of JT  have been suggested in refs.~\cite{Saad:2019lba,Post:2022dfi}.}. Further developments made explicit many detailed properties of the discrete spectrum (or, better, the ensemble of {\it candidate}  discrete spectra) for the first time~\cite{Johnson:2021zuo,Johnson:2022wsr}, from which many quantities can be computed directly, such as the spectral form factor and the quenched free energy. 

${\cal N}{=}2$ JT supergravity was still not well understood in matrix model terms until the recent work of Turiaci and Witten~\cite{Turiaci:2023jfa}, where  many features of the relevant random matrix model were proposed and successfully compared (perturbatively in topology) to features of the supergravity accessible through the Euclidean path integral. Among their crucial observations is the fact that matrix model descriptions  apply  in  {\it subsectors} of the overall model, due to the (perhaps surprising)   statistical independence of various subsectors from each other. Specifically, the leading (disc) order partition function for the  non-BPS\footnote{Here, BPS is short for Bogomol'nyi–Prasad–Sommerfield~\cite{Bogomolny:1975de,Prasad:1975kr}.} states is built from  spectral densities $\rho_{0,q}(E)$ for a multiplet with R-charge labelled by $q$,  given by~\cite{Stanford:2017thb,Mertens:2017mtv}:
\be
\label{eq:leading-spectral-density}
\rho_{0,q}(E)  = \frac{\e^{S_0}}{2\pi}\frac{\sinh(2\pi\sqrt{E-E_0(q)})}{4\pi^2 E}\Theta(E-E_0(q))\ ,
\ee
where $E_0(q){=}q^2/4{\hat q}^2$ is the minimum (threshold) energy in the sector (${\hat q}$ is the R-charge of the supercharge). When $|q|{<}{\hat q}$ there are BPS states present with  density: 
\be
\label{eq:BPS-density}
\rho_{\rm BPS}(E)=\frac{e^{S_0}}{4\pi^2}\sin\left(\frac{\pi q}{{\hat q}}\right)\delta(E)\ .
\ee
Ref.~\cite{Turiaci:2023jfa} proposed  that  in such a subsector, with~$\nu$  BPS states, there is a random matrix ensemble description with this spectral density, and it is of $(1{+}2\nu,2)$ type in the Altland-Zirnbauer (AZ)~\cite{Altland:1997zz} $(\boldsymbol{\alpha},\boldsymbol{\beta})$ classification. For vanishing $q$, $E_0(q){=}0$  and the leading density has  ${\sim}E^{-\frac12}$ ``hard edge'' behaviour. Otherwise, the tail of the  distribution is of the form ${\sim}\sqrt{E-E_0(q)}/E$. Several checks and arguments   supported this, including alignment with features of 
new topological recursion relations  generalizing those of Mirzakhani for the ${\cal N}{=}0$ case~\cite{Mirzakhani:2006fta,*Eynard:2007fi}, and those   in ref.~\cite{Stanford:2019vob}.

An important  next step is to find a natural non-perturbative definition of the matrix model that exhibits these features. This paper presents   one from which results can be efficiently extracted. Random matrix models can often be written in terms of families of orthogonal polynomials~\cite{Brezin:1978sv,Bessis:1980ss},  characterized by a recursion relation. Physical quantities are then expressed in terms of the polynomials. The recursion coefficients defining the polynomials  satisfy  difference equations that follow from the matrix model potential. In the double-scaling limit that yields   gravity, the difference  equations become  non-linear ordinary differential equations (ODEs), known (from the  literature of the early 1990s) as  ``string equations''\footnote{For a recent exposition in the JT gravity context, see the early sections of ref.~\cite{Johnson:2021tnl}.}. For a one-matrix model (the case of relevance here), only one such equation emerges in the limit. It will be shown below how to build   a string equation whose solutions capture key features of  the  non-BPS and BPS sectors of  ${\cal N}{=}2$ JT supergravity, yielding formulae~(\ref{eq:leading-spectral-density}) and~(\ref{eq:BPS-density}), and perturbative and non-perturbative physics beyond. The construction's components   are a family of multicritical models, a particular combination of which  yields the  model.

\section{The String Equation}
The %form of the 
central equation is for the function $u(x)$, where $x$ runs over the real line:
\be
\label{eq:big-string-equation}
u{\cal R}^2-\frac{\hbar^2}2{\cal R}{\cal R}^{\prime\prime}+\frac{\hbar^2}4({\cal R}^\prime)^2=\hbar^2\Gamma^2\ .
\ee
Here $\hbar{=}\e^{-S_0}$, and  ${\cal R}{\equiv}\sum_{k=1}^\infty t_k R_k[u]+x$ where the 
%$R_k[u]$ are polynomials (see below) in the function $u(x)$ and its $x$-derivatives. Also,~ . 
%The 
$ R_k[u]$ are the ``Gel'fand-Dikii''~\cite{Gelfand:1975rn} differential polynomials  in $u(x)$ and its derivatives, normalized here so that the non-derivative part has unit coefficient. They are of the form: $R_k{=}u^k+\cdots+\#u^{(2k-2)}$ where $u^{(m)}$ means the $m$th $x$-derivative. {\it e.g.}, $R_1{=}u$, $R_2{=}u^2{-}\frac{\hbar^2}{3}u^{\prime\prime}$, $R_3 {=} u^3+{\hbar^2}(u^\prime)^2/2+{\hbar^2}uu^{\prime\prime}+{\hbar^4}u^{(4)}/{10}$, {\it etc.} (Successive $R_k$ can be obtained by a recursion relation that won't be needed here.)

Equation~(\ref{eq:big-string-equation}) arose in early studies~\cite{Dalley:1991qg,Dalley:1991vr,Morris:1991cq,Dalley:1991xx,Dalley:1992br,Anderson:1991ku} of {\it positive} random matrix model ensembles for gravity applications (later identified to be of type  $(1{+}2\Gamma,2)$ in the AZ classification). The $N{\times}N$ matrices  of the model can be written as $M{=}Q^\dagger Q$ where $Q$ is a random $N{\times}(N{+}\Gamma)$ complex rectangular matrix. $M$ therefore has $\Gamma$ repeated zero eigenvalues. Such models can be given a polynomial potential $V(M)$, where the linear case is the classic Wishart model~\cite{10.2307/2331939}, whose low energy leading spectral density has the $\rho_0{\sim} E^{-\frac12}$ behaviour. Tuning coefficients $\{g_i\}$ in $V(M)$ to values $\{g^c_i\}$ such that there are $k$ additional  zeros at the edge, $\rho_0{\sim} E^{k-\frac12}$ yields~\cite{Dalley:1991qg} the $k$th  ``multicritical'' model. In the usual double scaling limit (taking $N{\to}\infty$ while zooming at the correct rate into the edge so as to access the universal physics there) the model is fully captured by the string equation~(\ref{eq:big-string-equation}) with ${\cal R}{\equiv}R_k[u]+x$. The function $u(x)$ is the part of the orthogonal polynomial recursion coefficients that emerge in the limit to capture the universal physics of the endpoint of the matrix model spectrum, and $\hbar$ is a renormalized~$1/N$ parameter, as is familiar in these settings. It is the topological expansion parameter. Solutions of  equation~(\ref{eq:big-string-equation}) are generically of the following form: $u(x){=}\sum_{g,n} \hbar^{2g+n}\Gamma^n u_{g,n}(x) +\cdots,$ where the combination $2g{+}n{=}2{-}\chi$, and $\chi$ is the Euler number of a surface with~$g$ handles and $n$ boundaries or punctures.  The ellipsis denotes non-perturbative physics. The perturbative expansion is asymptotic, and properties of such solutions have been explored in the early literature~\cite{Dalley:1991qg,Dalley:1992br,Johnson:2004ut}. In general, once the leading solution $u_{0,0}(x)$ is fixed (see below, where it is denoted $u_0(x)$ for simplicity), the higher order corrections $u_{g,n}(x)$ are determined recursively by substitution into the equation.  Most of the attention in this paper will be given to characterizing the class of leading solutions $u_0(x)$ needed to capture the physics of ${\cal N}{=}2$ JT supergravity, although some aspects of the non-perturbative physics defined by the model will be studied in Section~\ref{sec:non-perturb} and Appendix~\ref{app:non-perturb}.\footnote
{While some early attention had been given to the instanton effects implied in the asymptotic expansions (see {\it e.g.} refs.~\cite{Dalley:1991qg,Johnson:2004ut}) of the string equation, attributing them to one-eigenvalue/D-brane non-perturbative effects in the spirit of early refs.~\cite{David:1990ge,*Shenker:1990uf},  it would be interesting to explore to what extent the more recent resurgence and transseries techniques developed for random matrix models and JT gravity (see {\it e.g.} refs.~\cite{Marino:2007te,*Marino:2008ya,*Gregori:2021tvs,*Eynard:2023qdr,*Schiappa:2023ned,*Vega:2023mpu}) can capture some of the non-perturbative physics seen in the direct study (analytic and numerical) of the full solutions. It would be especially interesting in the case of the class of solutions that will later be the focus of this paper, where the limit $\Gamma{\to}\infty$ as $\hbar{\to}0$ (while holding the product fixed) will be key.}  

The string equation resulting from the most general polynomial potential can be written using the more general form for ${\cal R}$ given below~(\ref{eq:big-string-equation}).
 A particular~$t_k$ determines how much the multicritical model labelled by~$k$ contributes. Their values  will be fixed shortly by matching to disc order.

In the random matrix model description  given above,~$\Gamma$ is naturally to be thought of as an integer, but studies over the years have shown that solutions of Eq.~(\ref{eq:big-string-equation}) with more general $\Gamma$ have physical relevance too.\footnote{For an early example, the case of $\frac12$-integer $\Gamma$  was recognized to be extremely natural in the light of the  discovery~\cite{Dalley:1992br} of a map from solutions of Eq. (\ref{eq:big-string-equation}) to solutions of a Painlev'e~II hierarchy of equations characterized by a constant $C{=}\frac12\pm\Gamma$, where integer $C$ counts the numbers of flavours of quark in  multicritical generalizations~\cite{Periwal:1990gf,*Periwal:1990qb,*Minahan:1991ry,*Minahan:1991pv,*Gross:1991aj,*Gross:1991ji} of the Gross-Witten-Wadia model~\cite{Gross:1980he,*Wadia:2012fr}.}   Later in this paper it will be confirmed that $\Gamma$  will be identified with the number of BPS states within an ${\cal N}{=}2$ multiplet labelled by $q$ (a number denoted as $\nu$ or $\rho_{\rm BPS}$ in the Introduction).  Remarkably, this identification will not be a choice but a {\it condition}  arising from  internal consistency of the model if the solutions constructed are to capture the correct non-BPS physics.

 Once boundary conditions are specified (see below), the string equation~(\ref{eq:big-string-equation}) determines  $u(x)$,  the continuum limit of the recursion coefficients determining the (double scaled) orthogonal polynomials, denoted $\psi(x,E)$. They are functions of the scaled eigenvalue, denoted $E$, and labelled by a continuous index~$x$. Rather nicely for this class of models, $\psi(E,x)$ are determined (up to  normalization) as wavefunctions of a Schr\"odinger problem: 
\be 
\label{eq:schrodinger-problem}
\left[-\hbar^2\frac{\partial^2}{\partial x^2}+u(x)\right]\psi(x,E)=E\psi(x,E)\ ,
\ee
for which $u(x)$ is the potential. The spectral density of the model is given in terms of $\psi(E,x)$ as follows:
\be
\label{eq:spectral-density-exact}
\rho(E) =\int_{-\infty}^\mu \left|\psi(x,E)\right|^2 dx\ , 
\ee
and the value of  $\mu$ will be determined below by matching to disc level physics. 
%\footnote{Before the continuum limit, there are $N$ orthogonal polynomials needed to capture all the physics, and the density  results from a sum over an index $n$ up to $N$. After going to the continuum and taking the scaling limit, $\mu$ emerges as the scaled portion of the uppermost value of the index.}
The disc order spectral density arises from taking the leading Wentzel–Kramers–Brillouin (WKB) form of the wavefunctions, giving:
\be
\label{eq:spectral-density-leading}
\rho_{0}(E) \!= \frac{1}{2\pi\hbar}\int_{-\infty}^\mu\!\frac{\Theta(E{-}u_0(x)) dx}{\sqrt{E-u_0(x)}}=\frac{1}{2\pi\hbar}\int_{E_0}^E\!\frac{f(u_0)du_0}{\sqrt{E-u_0}} \ ,
\ee
where $u_0(x)$ is the leading perturbative piece of $u(x)$ obtained by sending $\hbar{\to}0$.  
In the second expression $E_0{=}u_0(\mu)$, and $f(u_0){=}{-}\partial x/\partial u_0$.

\section{$E_0{=}0$ multicritical Model Decomposition}
This case is treated in a manner similar to what has been done for earlier matrix model studies of ordinary JT and various ${\cal N}{=}1$ JT models\cite{Johnson:2019eik,Johnson:2020mwi,Johnson:2020exp}. The multicritical models in question~\footnote{These models were discovered and extensively studied long ago in refs.~\cite{Morris:1991cq,Dalley:1992qg,Dalley:1992vr,Johnson:1992wr}, and later identified in ref.~\cite{Klebanov:2003wg} as being of type ``0A'' in the minimal string context. } are specified by boundary conditions that can be written for $\hbar\to0$, where the string equation becomes $u_0{\cal R}_0^2=0$, with ${\cal R}_0=\sum_k t_k u_0^k+x$: 
\bea 
u_0=0 \,\,\, \mbox{as}\,\,\, x\to+\infty\ , \,\,\mbox{and} \,\,
 {\cal R}_0=0 \,\,\, \mbox{as}\,\,\, x\to-\infty\ .\label{eq:boundary-conditions}
 \eea
 Next, the $t_k$ are uniquely determined by requiring that the leading (non-BPS) spectral density comes out as equation~(\ref{eq:leading-spectral-density}) (with $E_0{=}0$ for far). It is easiest to simply expand $\rho_{0,q}(E)$ as a power series in $E^\frac12$, and use the fact that for the $k$th model, the string equation in the negative~$x$ region is $u_0^k{+}x{=}0$, so $f(u_0){=}ku_0^{k-1}$, and  integral~(\ref{eq:spectral-density-leading}) gives~\cite{Johnson:2020heh} $\rho_0^{(k)}(E){=}C_kE^{k-\frac12}/2\pi\hbar$, where $C_k{=}2^{2k{-}1}k ((k{-}1)!)^2/(2k{-}1)!$. Matching coefficients for the full expansion gives a new equation for the multicritical model ingredients for ${\cal N}{=}2$ supergravity (for $E_0{=}0$):
\be
\label{eq:new-teekay}
t_k=\frac{\pi^{2k-1}}{2(2k+1)(k!)^2}\ ,\quad{\rm with}  \quad\mu{=}(2\pi)^{-1}\ .
\ee
 It is crucial to notice (as first observed in the studies of ${\cal N}{=}1$ cases in ref.~\cite{Johnson:2020heh}) that the key $1/\sqrt{E}$ contribution to the  tail of the leading spectral density comes from the fact that $u_0{=}0$ in the positive $x$ region, and so carrying out the part of  integral~(\ref{eq:spectral-density-leading}) there gives a contribution $(2\pi\hbar)^{-1}\mu/\sqrt{E}$, which also fixes the value of $\mu$.

The equation for the $t_k$ can be used to write the leading order  piece of the string equation for $u_0(x)$ in a closed form, since ${\cal R}_0=\sum_k t_k u_0^k+x=0$ becomes:
\bea
x&=&\frac{1}{4\pi}\left[\pi I_1(2\pi\sqrt{u_0})L_0(2\pi\sqrt{u_0})\right.\\
&&-\left.\pi I_0(2\pi\sqrt{u_0})L_1(2\pi\sqrt{u_0}) -2I_0(2\pi\sqrt{u_0})+2 \right]\ ,\nonumber
\eea
where $I_n(y)$ and $L_n(y)$ are, respectively,  the modified Bessel and Struve function in $y$ of order~$n$.
(This generalizes the case for ${\cal N}{=}1$ models derived in refs.~\cite{Johnson:2020heh,Johnson:2020exp}.) 

\section{$E_0{\neq}0$ multicritical Model Decomposition}
For this case, a new approach is needed in order to get the $\sqrt{E-E_0}/E$ edge behaviour. Some experimentation shows that $E_0$-dependent parameters $\{{\tilde t}_k(E_0),{\tilde\mu}(E_0)\}$ must be sought, giving a new multicritical model recipe. By way of preparation for such a decomposition, a  generalization of   $C_k$ mentioned above equation~(\ref{eq:new-teekay}) for integrating $f(u_0){=}ku_0^{k-1}$ for non-zero~$E_0$ is needed. The derivation is  given in Appendix~\ref{app:cee_kay_extended}, with the result  $\rho_0^{(k)}(E_0,E)= {\widetilde C}_k(E_0,E)E^{k-\frac12}/2\pi\hbar$, where:
\bea
\label{eq:new_Cee_kay}
{\widetilde C}_k(E_0,E)=&&\\ \nonumber
&&\hskip-1.5cm 
\frac{2k}{2^{2k-2}}\sum_{i=1}^k\frac{(2k-1)!(-1)^{i-1}}{(k-i)!(k+i-1)!}\frac{\cos\left[(2i-1)\theta_0\right]}{(2i-1)}\ ,
\eea
with $\cos\theta_0{=}\sqrt{1-{E_0}/{E}}$. (The first few cases are displayed explicitly in equations~(\ref{eq:examples-of-minimals}).)
Next  is to discover how to expand the leading spectral density~(\ref{eq:leading-spectral-density}) into individual  components that allow a determination of~${\tilde t}_k(E_0)$. The strategy is to  expand $(E-E_0)^\frac12{=}E^\frac12(1-E_0/2E+\cdots)$ and then expand the $\sinh$ function that it sits inside. This results in a  complicated expression of the  form:
\be
\label{eq:expansion}
\hskip-0.1cm \rho_{0,q}(E,E_0)\simeq\sum_{k\geq1} A_k(E_0) E^{k-\frac12}+\sum_{k\geq1} B_k(E_0) E^{\frac12-k}\ ,
\ee
an $E_0$-dependent mix of fractional powers of $E$, including not just positive powers but also  negative powers {\it beyond} $E^{-\frac12}$, in stark contrast to the $E_0$ case. For example: 
\be
B_1(E_0)=\frac{1}{4\pi^2}-\frac{E_0}{4}+\frac{\pi^{2} E_0^{2}}{16}
-\frac{\pi^{4} E_0^{3}}{144}+\cdots\ ,
\ee the first term giving the $E^{-\frac12}$ behaviour that is present in the  $E_0{=}0$ case, but now in addition there are  $E_0$-dependent terms at this order and also at more negative  powers of $E$.  Such terms are at first puzzling, but they are  the  contributions needed to cancel  the hard edge behaviour of the tail of $\rho(E)$ in order to truncate it at~$E_0$. 

In fact, examining~(\ref{eq:new_Cee_kay}) for the  behaviour resulting from using a multicritical model shows that (after also expanding in small~$E_0$) such negative fractional powers of $E$ are indeed  possible. The issue to hand is whether they can be made to match what is present in equation~(\ref{eq:expansion}). A tentative decomposition into multicritical models is given by the following tree level string equation for the $x{<}0$ region, generalizing~(\ref{eq:boundary-conditions}):
%\be
%\label{eq:new-string-equation}
${\cal R}_0{\equiv}\sum_{k=1}^\infty {\tilde t}_k(E_0)u_0^k+x=0.$  %\ee
The form of the resulting spectral density can be determined by putting $f(u_0)=\partial_{u_0}{ {\cal R}}_0$ 
%= \sum_k k{\tilde t}(E_0) u_0^{k-1}$ 
into equation~(\ref{eq:spectral-density-leading}) and expanding in~$E_0/E$. As before, comparing powers of $E^{k-\frac{1}{2}}$ for positive $k$  gives the ${\tilde t}_k(E_0)$, at least order by order in $E_0$. For example, to quadratic order, the first few ${\tilde t}_k(E_0)$ are: %can be written as:
\bea
{\tilde t}_1(E_0) &=& t_1\left(1-\frac{1}{2} \pi^{2} E_0 +\frac{1}{12}\pi^4E_0^2+\cdots\right)\ ,\nonumber\\
{\tilde t}_2(E_0) &=& t_2\left(1-\frac{1}{3} \pi^{2} E_0 +\frac{1}{24} \pi^{4}E_0^2+\cdots\right)\ , \nonumber\\
{\tilde t}_3(E_0) &=& t_3\left(1-\frac{1}{4} \pi^{2} E_0 +\frac{1}{40}\pi^4E_0^2+\cdots\right)\ ,
\eea
where $t_1{=}\pi/6$, $t_2{=}\pi^3/40,$ and $t_3{=}\pi^5/504$ were determined earlier in~(\ref{eq:new-teekay}).
This still leaves an infinite number of terms (the $B_k$) to match. One might hope these are automatically fixed to be the negative powers that come with the %multicritical model 
expression~(\ref{eq:new_Cee_kay}) for ${\widetilde C}_k(E_0,E)$, but this is not borne out, and an infinite number of terms fail to match. 
%So it seems  that the enterprise is doomed, since there are no multicritical models that can adjust the missing negative powers  once the ${\tilde t}_k(E_0)$ are fixed.

Salvation comes from  the $x{>}0$ region. What is needed is a natural generalization of the $u_0(x){=}0$ solution used for $E_0{=}0$.  Another kind of special classical solution to the string equation~(\ref{eq:big-string-equation}) can be obtained by sending $\Gamma{\to}\infty$ while $\hbar{\to}0$, holding fixed  $\Gamma^2\hbar^2={\tilde\mu}^2E_0$. Drop all other terms involving $\hbar$ as usual, and turning off all the $t_k$s leaves ${\cal R
}_0{=}x$, giving:
\be
\label{eq:rational}
u_0(x)=\frac{{\tilde\mu}^2E_0}{x^2}\ .
\ee
(Note that $u_0({\tilde\mu}){=}E_0$ for this solution.) Inserting this  into WKB integral~(\ref{eq:spectral-density-leading}) for $x{>}0$  (using lower limit $x{=}{\tilde\mu}\sqrt{E_0/E}$ and upper limit $x{=}\tilde\mu$), yields $\frac{\tilde\mu}{2\pi}\sqrt{E-E_0}/E$.  Expanding this produces  new negative fractional powers of $E$. Their coefficients are determined by writing  ${\tilde\mu}(E_0)$ as a power series in $E_0$, %(with leading term  $1/2\pi$)  
fixing each term's coefficient to match the remaining terms in $B_0$ ({\it i.e.} order $E^{-\frac12}$). The first few terms  are:
\be
\label{eq:first-few}
{\tilde\mu}=\frac{1}{2\pi}\left(1-\frac23\pi^2E_0+\frac{2}{15}\pi^4E_0^2-\frac{4}{315}\pi^6E_0^3+\cdots\right)
\ee
Checking the same orders in $B_2$ (order $E^{-\frac32}$) shows that those are matched too, and so on to increasingly negative powers! Remarkably, the procedure has succeeded in reproducing the disc spectral density~(\ref{eq:leading-spectral-density}) using classical string equation ingredients, at least perturbatively in~$E_0$. Actually, there is another miracle afoot that will reveal exact expressions in $E_0$,  arising next in the BPS sector.

\section{The BPS sector}
The solution~(\ref{eq:rational}) related the parameter $\Gamma$, the number of zero energy matrix model states,  to $E_0$ through (after taking the positive  square root) $\Gamma{=} \hbar^{-1}{\tilde\mu}\sqrt{E_0}$. The coefficients of the  $E_0$ expansion~(\ref{eq:first-few}) of ${\tilde \mu}$ discovered  above are suggestive, and exploring higher orders shows that it is simply the expansion of: 
\be
\label{eq:mu-result}
{\tilde \mu} = \frac{1}{4\pi^2\sqrt{E_0}}\sin(2\pi\sqrt{E_0})\ ,
\ee 
but since $\hbar{=}e^{-S_0}$ and $E_0{=}q^2/4{\hat q}^2$, this results in the exact result 
$\Gamma {=} %\pm
\frac{e^{S_0}}{4\pi^2} \sin\left({\pi q}/{\hat q}\right)$,
%\ee 
  {\it precisely} the content of  equation~(\ref{eq:BPS-density})  for the number of BPS states!~\footnote{Note that for multiplets with $|q|\geq{\hat q}$, there are no BPS states. The edge of the BPS regime is just where $\Gamma$ starts becoming  negative. While the string equation can handle solutions with either sign of $\Gamma$, it  would be interesting to understand if negative $\Gamma$ has a physical interpretation in the current context. In the BPS range, for small values of the parameters, the string equation seems well-behaved, but there is the possibility that non-perturbative features come into play for the wider range of solutions. This is worth further exploration.}
This remarkable outcome suggests that the infinite set of perturbative (in~$E_0$) expressions for ${\tilde t}_k(E_0)$ can be also re-summed into something simpler. Some experimentation confirms this, resulting in   the following satisfying expression:
\bea
\label{eq:new_teekay_exact}
{\tilde t}_k(E_0) &=& t_k %\cdot 
\frac{2^k k!}{(2\pi\sqrt{E_0})^k}%\cdot 
J_k(2\pi\sqrt{E_0}) 
%\\&=&   \frac{ k!}{2(2k+1)(\pi\sqrt{E_0})^k}\cdot J_k(2\pi\sqrt{E_0}) 
\ .
\eea
%where 
$J_k(y)$
is the $k$th Bessel function of $y$, and $t_k$ is given in~(\ref{eq:new-teekay}).

This completes the decomposition into multicritical models, fixing the content and structure of the the full non-linear string equation~(\ref{eq:big-string-equation}), defining the model 
%. Perturbative  solutions of the string equation seeded by the classical content will yield results equivalent to those of the perturbation theory of ref.~\cite{Turiaci:2023jfa}, (see ref.~\cite{Johnson:2021owr} for aspects of an ${\cal N}{=}1$ case) but of course this setting  provides  non-perturbative solutions that allows for explorations of the matrix model (and hence  gravity) 
well beyond the topological expansion. 

\section{Non-Perturbative Results}
\label{sec:non-perturb}
\begin{figure}[b]
%\begin{wrapfigure}{r}{0.45\textwidth}
\centering
\includegraphics[width=0.48\textwidth]{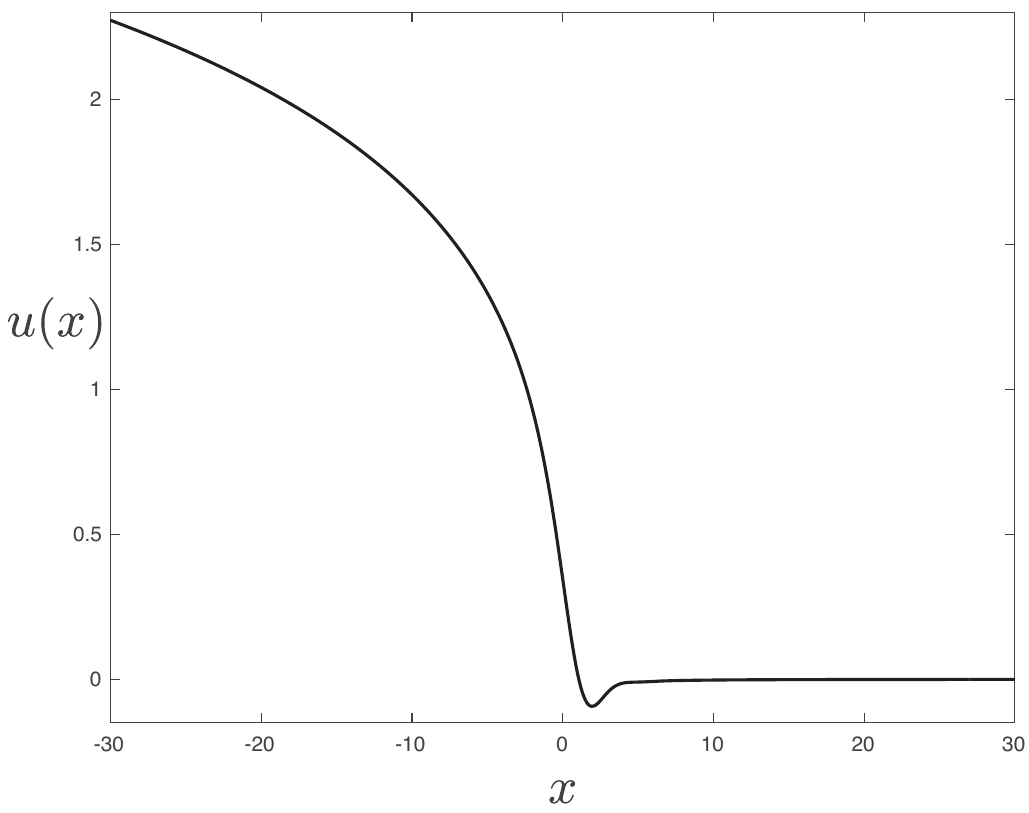}
%:
\caption{\label{fig:u-solutions} The solution~$u(x)$ for $E_0{=}0$.  Here $\hbar{=}{\rm e}^{-S_0}{=}1$. }
%\end{wrapfigure}
\end{figure}

The full non-linear  equation~(\ref{eq:big-string-equation}), with the $t_k$ of equation~(\ref{eq:new-teekay}) (or ${\tilde t}_k(E_0)$ of~(\ref{eq:new_teekay_exact}), with $\hbar\Gamma{=}{\tilde\mu}\sqrt{E_0}$), defines a solution for $u(x)$, given the boundary conditions~(\ref{eq:boundary-conditions}). Many earlier studies suggest that smooth unique solutions exist for a wide range of parameters~\cite{Johnson:1992pu}. A concern might be that it is formally an infinite order equation (since all $t_k$ are involved), but note~\cite{Johnson:2020exp} that a sensible $u(x)$ still exists by  virtue of the fact that the $t_k$ rapidly fall in magnitude at higher~$k$, and hence the solution's modifications from higher derivatives are vanishingly small at  higher order. This suggests a truncation scheme whereby solving up to some high enough order suffices to capture (to any desired accuracy) the physics. This has proven very successful for previous work of this kind on ${\cal N}{=}0$ and ${\cal N}{=}1$ JT gravity~\cite{Johnson:2020exp,Johnson:2022wsr}. A truncation at $k{=}7$ was chosen, resulting in the need to numerically solve the non-linear ODE at 15th order~\footnote{The basic ODE is order $2k$, but taking an additional derivative of the whole equation reduces the non-linearity. See refs.~\cite{Johnson:2020exp,Johnson:2022wsr} for a guide to the numerical methods.}. For the case $E_0{=}0$  the resulting $u(x)$ is shown in figure~\ref{fig:u-solutions}, with $\hbar{=}1$ for illustration.
\begin{figure}[h]
%\begin{wrapfigure}{r}{0.45\textwidth}
\centering
\includegraphics[width=0.48\textwidth]{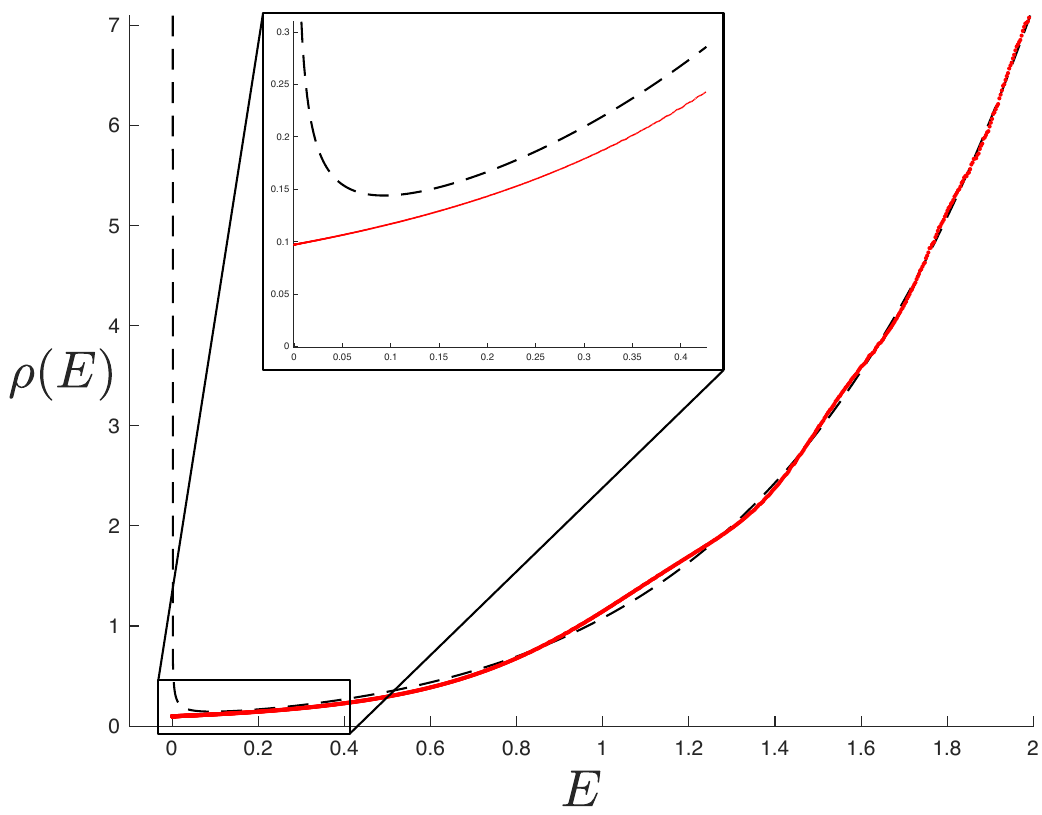}
%:
\caption{\label{fig:spectral-density} The solid (red) curve is the non-perturbative spectral density $\rho(E)$ for an $E_0{=}0$ non-BPS multiplet. The dashed line shows $\rho_0(E)$. Inset: Close-up of the endpoint, showing a finite non-zero $\rho(E{=}0)$ (see text). Here $\hbar{=}{\rm e}^{-S_0}{=}1$. }
%\end{wrapfigure}
\end{figure}
The next step is to solve system~(\ref{eq:schrodinger-problem}) for the  $\psi(x,E)$, using~$u(x)$ as a potential, and finally to construct the spectral density according to the integral~(\ref{eq:spectral-density-exact}), using ${\tilde\mu}$ from equation~(\ref{eq:mu-result}) as the upper limit (it is in the range ($-1/3\pi^2$,$1/2\pi$),  depending on~$E_0$). The resulting $\rho(E)$ for $E_0{=}0$ is shown in figure~\ref{fig:spectral-density}, with the dashed line showing the disc result (with $\hbar{=}1$). % (A figure showing $\rho(E)$ for $E_0{=}\frac14$ is given and discussed in the Supplemental Material.) 
Non-perturbative corrections produce undulations about the dashed line, becoming increasingly marked at lower~$E$, like the ordinary and ${\cal N}{=}1$ JT cases~\cite{Johnson:2020exp}. The corrections are extremely important at the lowest energies, and for the $E_0{=}0$ case,  the  ${\sim}E^{-\frac12}$ divergence of the leading disc result  is rendered  finite. (This  finiteness could have been anticipated\cite{Johnson:2021rsh} since  the tail is essentially the (1,2) Bessel model. Also, just as in that case, it is important to note that despite appearances, the spectral density does {\it not} continue smoothly to some $E{<}0$ regime. The matrix model's spectrum is positive by construction.) 

It is also possible to construct solutions for $E_0{\neq}0$, and an example is given in Appendix~\ref{app:non-perturb}. It is of interest to also  study solutions with large numbers of BPS states ({\it i.e.,} scaling with ${\rm e}^{S_0}$ for large $S_0$), but this  requires solving for $u(x)$ at small~$\hbar$, which has so far not yielded success: $u(x)$ now varies more quickly  with~$x$ in the central region, presenting a more difficult numerical challenge than for $\hbar{\sim}1$ where it is more smooth. This is left for future work.

\section{Closing Remarks}
This paper provides a fully unified and non-perturbative definition of the double-scaled random matrix model of ${\cal N}{=}2$ JT supergravity discussed perturbatively  by Turiaci and Witten in ref.~\cite{Turiaci:2023jfa}. Various features and properties of the model that were deduced using a combination of   considerations there (gravity  path integral, loop equations, topological recursion, {\it etc.})  are seen to emerge directly from the definition given here, embodied in properties of a special class of solutions of the string equation~(\ref{eq:big-string-equation}).

As  in previous JT gravity examples, this definition allows exploration of  a great deal of non-perturbative quantum gravity physics. This includes features that  require knowledge of the underlying discrete spectrum (the %(quenched) 
free energy, the spectral form factor, {\it etc.}). Using a Wignerian~\cite{10.2307/1970079} perspective (emphasized in refs.~\cite{Johnson:2022wsr,Johnson:2022hrj}),  random matrix models of gravity are  understood non-perturbatively simply as tools for studying the statistics of all candidate holographic discrete spectra that are consistent with the leading disc data. 
Powerful tools (such as Fredholm determinants, introduced in this context in ref.~\cite{Johnson:2021zuo}), can now yield detailed statistical aspects of the spectra.  (Appendix~\ref{app:peaks} shows the probability distribution peaks for the first five energy levels for the $E_0{=}0$ case.) 

Notably, $\Gamma$ BPS states were extremely natural to incorporate into the model because it is an ensemble of matrices with some fixed number ($\Gamma$) of zeros. The zeros naturally generate a gap in the spectrum  that grows with~$\Gamma$ (this can also be seen  in toy Bessel models~\cite{Johnson:2021rsh}). This is very reminiscent of properties of  extremal BPS  black holes in higher dimensions~\cite{Preskill:1991tb,Heydeman:2020hhw}, and in this ${\cal N}{=}2$ context, would seem to be a precise non-perturbative model of the phenomenon. This is clearly worth further study.

%\begin{acknowledgments}
\section*{Acknowledgments}
CVJ  thanks  the  US Department of Energy (\protect{DE-SC} 0011687) and the Aspen Center for Physics (under NSF grant PHY-2210452) for  support, Felipe Rosso, Joaquin Turiaci, and Edward Witten for remarks, and  Amelia for her support and patience.    
%\end{acknowledgments}

%\vfill\eject
%\newpage

\appendix

\begin{widetext}

\section{A generalization of $C_k$}
\label{app:cee_kay_extended}
This is a derivation of a generalization of the formula for $C_k$ (shown just above~(\ref{eq:new-teekay}), that includes the effect of the lower bound $E_0$ on the integral~(\ref{eq:spectral-density-leading}) for an individual multicritical model: 
\be\rho_0^{(k)}(E_0,E)={\widetilde C}_k(E_0,E)E^{k-\frac12}/2\pi\hbar\ .
\ee The route is  a straightforward continuation of what was already outlined in ref.~\cite{Johnson:2020heh}. After a change of variables to $z{=}u_0/E$, all $E$ dependence factored out, leaving an integral over $z$ from~$0$ to~$1$. The messy $(1{-}z)^{-\frac12}$ part of the integrand can be eliminated by changing variables to $z{=}\sin^2\theta$ after which the problem is simply the $\theta$-integral of $(\sin\theta)^{2k-1}$ from $0$ to $\frac{\pi}{2}$. This was  performed by deriving a useful identity in terms of sums of $\sin(n\theta)$ for  odd $n=1,3,\cdots,2k{-}1$. After integration, the only contribution came from evaluating $\cos(n\theta)$ at the lower ($\theta{=}0$) limit. In the present case however, the lower limit is $\theta_0{=}\sin^{-1}\sqrt{{E_0}/{E}}$, which introduces the non-trivial dependence through $\cos(n\theta_0)$, straightforwardly rewritten for any $n$ in terms of $\cos\theta_0{=}\sqrt{1-{E_0}/{E}}$. The key formulae to report then are the identity (derived by exponentiating and expanding  $\sin\theta{=}\left(\e^{i\theta}{-}\e^{-i\theta}\right)\!/2i$):
\begin{equation}
(\sin\theta)^{2k-1}=\frac{1}{2^{2k-2}}\sum_{i=1}^k\frac{(2k-1)!(-1)^{i-1}}{(k-i)!(k+i-1)!}\sin\left[(2i-1)\theta\right]\ ,
\end{equation}
 and upon integration and including all factors:
\bea
{\widetilde C}_k(E_0,E)= \frac{2k}{2^{2k-2}}\sum_{i=1}^k\frac{(2k-1)!(-1)^{i-1}}{(k-i)!(k+i-1)!}\frac{\cos\left[(2i-1)\theta_0\right]}{(2i-1)}\ ,
\eea
which when $\theta_0$ vanishes, re-sums to produce  the $C_k{=}2^{2k{-}1}k ((k{-}1)!)^2/(2k{-}1)!$ used in the main text.
For example the first few are:
\bea
\label{eq:examples-of-minimals}
\rho_0^{(1)}\! &=& \!2\frac{(E-E_0)^\frac12}{2\pi\hbar}\ ,\,\, \rho_0^{(2)} = \frac83\frac{(E-E_0)^\frac12}{2\pi\hbar}\left(E+\frac12E_0\right)\ ,\nonumber\\
\rho_0^{(3)}\! &=&\! \frac{16}{5}\frac{(E-E_0)^\frac12}{2\pi\hbar}\left(E^2+\frac12EE_0+\frac38E_0^2\right)\ .
\eea

\section{Non-perturbative results for $E_0{\neq}0$}
\label{app:non-perturb}
As mentioned in the main part of the text, it is also possible to explore $E_0{\neq}0$, and have some non-zero number of BPS states present. Curves for $u(x)$ and $\rho(E)$ are given in figure~\ref{fig:spectral-density-E0} for the case $E_0{=}\frac{9}{49}, \hbar{=}1$. 
As discussed earlier for the $E_0{=}0$ case, despite appearances, the spectral density does
{\it not} continue smoothly to some E<0 regime. The matrix model has positive eigenvalues by construction. The observed behaviour (the spectrum stopping at some finite density at $E{=}0$) is familiar from simpler exact models of positive matrices such as the (1,2) (Bessel) model.

It is to be expected that there is some non-perturbative incursion of states into the gap (due to quantum tunneling of the wavefunctions $\psi(E,x)$ into the region of the integration in equation~(\ref{eq:spectral-density-exact})), but for  $\hbar{=}1$, the effects are very strong, as can be seen in the figure (see inset). Moreover, for this value of $\hbar({=}{\rm e}^{-S_0})$ one cannot get a large number of BPS states such that $\Gamma{\sim}{\rm e}^{S_0}$, so this region of parameter space is perhaps not as interesting as for small $\hbar$. As mentioned in the main text, however, small $\hbar$ values are harder to control numerically (so far) and so further results for the $E_0{\neq}0$ sector must await further work.

\begin{figure}[t]
%\begin{wrapfigure}{r}{0.45\textwidth}
\centering
\includegraphics[width=0.48\textwidth]
{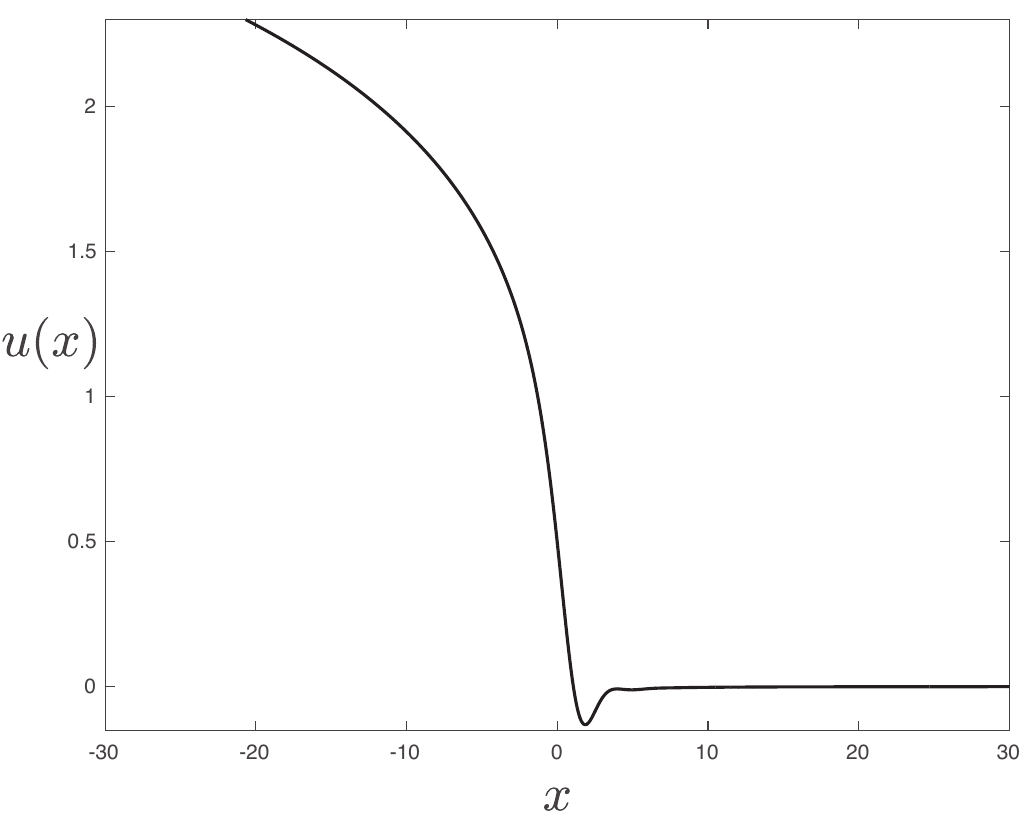}
\includegraphics[width=0.48\textwidth]
{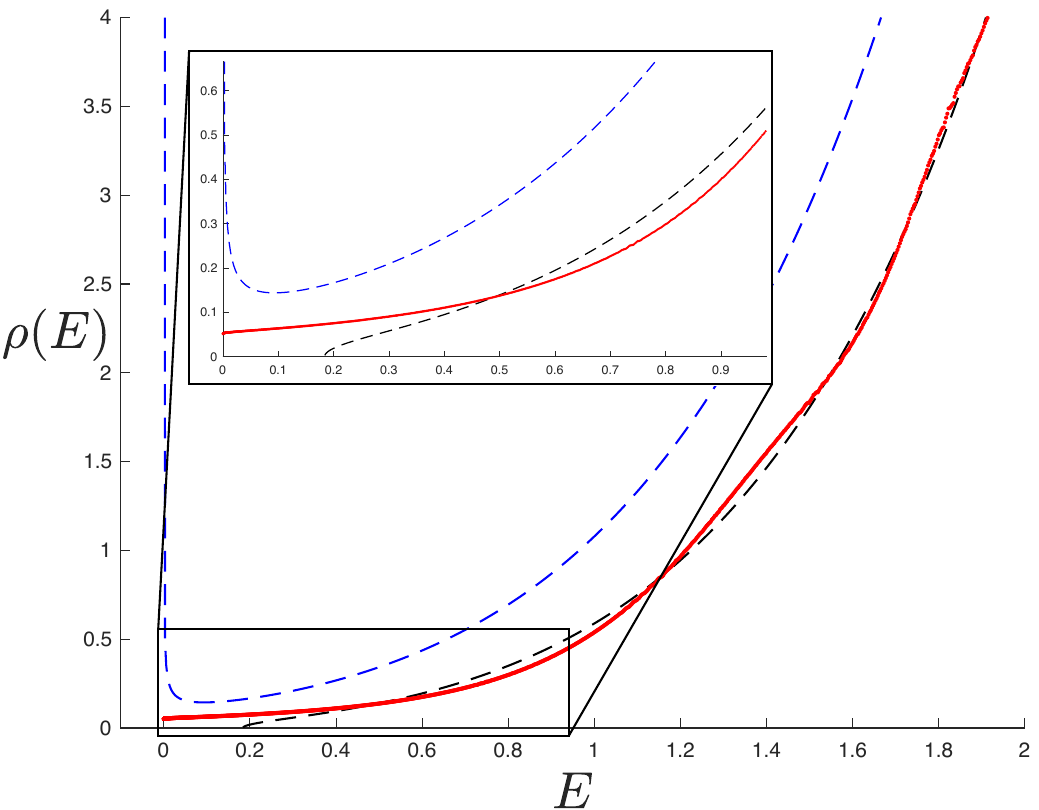}
\caption{\label{fig:spectral-density-E0} On the left is the non-perturbative potential $u(x)$  and on the right, the solid (red) curve is the  resulting spectral density  (non-BPS) for  $E_0{=}\frac{9}{49}$. (The lower dashed curve is $\rho_0(E)$ for  $E_0{=}\frac{9}{49}$ and the upper is $\rho_0(E)$ for $E_0{=}0$.) Inset: Close-up of the endpoint of the distribution.  Here $\hbar{=}{\rm e}^{-S_0}{=}1$. }
%\end{wrapfigure}
\end{figure}

\section{Underlying probability peaks}
\label{app:peaks}
As mentioned  in the main text, one of the main benefits of having a fully non-perturbative definition of the random matrix model (in matrix model terms) is that  it gives direct access to the microscopic details of the underlying ensemble of spectra that pertain to the gravity in a holographic sense.  This allows for a precise Wignerian interpretation of this kind of matrix model of gravity as simply the exploration of the ensemble of all the dual quantum gravity spectra that are consistent with the input data, which is the leading disc-level information. (This is spelled out in refs.~\cite{Johnson:2022wsr,Johnson:2022hrj}.) Specifically, the fully non-perturbative  matrix model spectral density is simply a sum $\rho(E){=}\sum_{n=0}^\infty p(n;E)$, where   $p(n;s)$ is the probability distribution of the $n$th energy level, as a function of energy $s$. This is an under-explored alternative (or dual) language that is entirely complementary to the procedure of perturbatively expanding the Euclidean 2D gravity path integral, and reaches well beyond it. As an example, the $p(n;s)$ were computed (using Fredholm determinants constructed from $\psi(E,x)$~\cite{Johnson:2021zuo,Johnson:2022wsr}) for the model presented in the main text, for the $E_0{=}0$ non-BPS sector, and the result presented in figure~\ref{fig:fredholm-peaks-E0-zero} (picking $\hbar{=}1$, and $\Gamma{=}0$). The disc contribution $\rho_0(E)$, full density $\rho(E)$, and (parts of) the five peaks are shown.  It is possible to do this for the $E_0{\neq}0$ non-BPS cases too. It should also be possible to derive an ODE describing the first peak representing the distribution of the lowest-lying non-BPS states, as done recently in this gravity context in ref.~\cite{Johnson:2022pou} for ordinary JT gravity. (Such a distribution gives a generalization of the well-known Tracy-Widom description of the distribution of lowest states the Gaussian Hermitian matrix model~\cite{Tracy:1992rf}.) The  results of such explorations  will be reported upon more fully in later work.

\begin{figure}[h]
%\begin{wrapfigure}{r}{0.45\textwidth}
\centering
\includegraphics[width=0.45\textwidth]{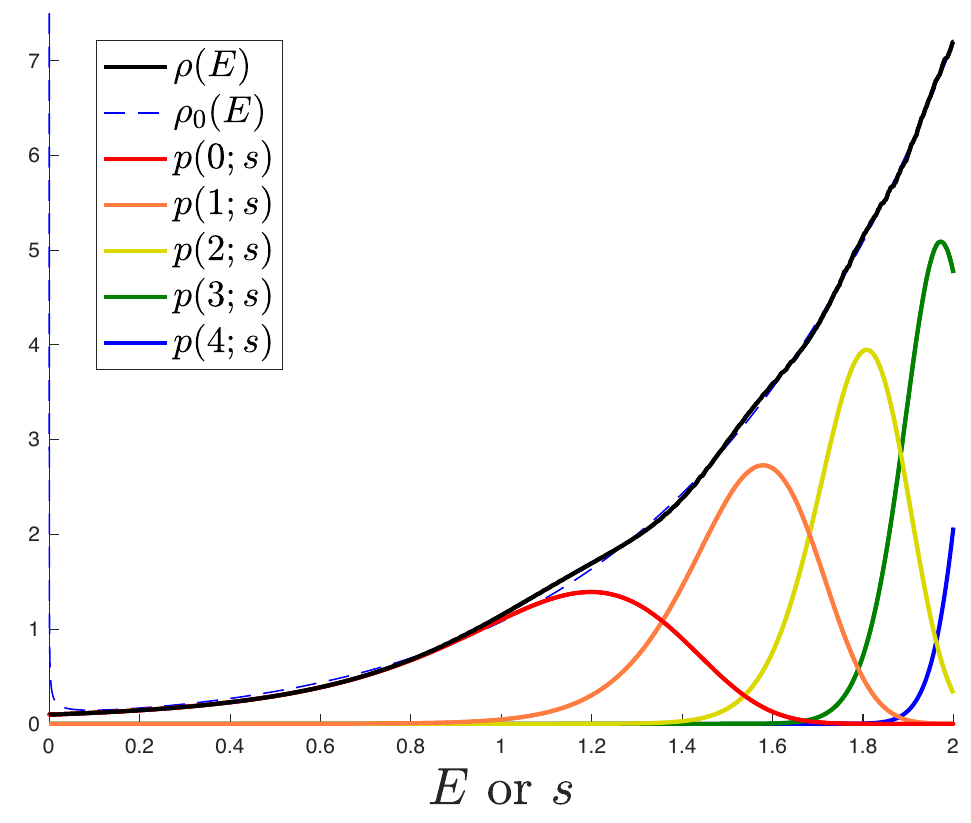}
%:
\caption{\label{fig:fredholm-peaks-E0-zero} The probability distribution peaks $p(n;s)$, for energy $s$, of the first five states ($n{=}0,1,\hdots$) in the $E_0{=}0$ non-BPS sector for $\Gamma{=}0$. See the legend in the inset for labelling. Here $\hbar{=}{\rm e}^{-S_0}{=}1$.}
%\end{wrapfigure}
\end{figure}

\end{widetext}

\bibliographystyle{apsrev4-1}
\bibliography{Fredholm_super_JT_gravity1,Fredholm_super_JT_gravity2}

\end{document}